\documentclass[a4paper,11pt]{article}

\usepackage{amsmath,amssymb,amsfonts,indentfirst,graphicx,color,anysize}

\def\bea{\begin{eqnarray}}
\def\eea{\end{eqnarray}}

\def\pp{\mbox{$p$-$p$}}

\def\pa{\mbox{$p$-A}}

\def\pbpb{\mbox{Pb-Pb}}
\def\ppb{\mbox{$p$-Pb}}
\def\pw{\mbox{$p$-W}}
\def\pbe{\mbox{$p$-Be}}

\def\pn{\mbox{$p$-N}}

\def\nn{\mbox{N-N}}

\def\pt{$p_t$}
\def\mt{$m_t$}
\def\yt{$y_t$}
\def\yz{$y_z$}

\def\nch{$n_{ch}$}
\def\mmpt{$\bar p_t$}

\marginsize{3cm}{3cm}{2cm}{2cm}
\title{
	\includegraphics[width=0.35\textwidth]{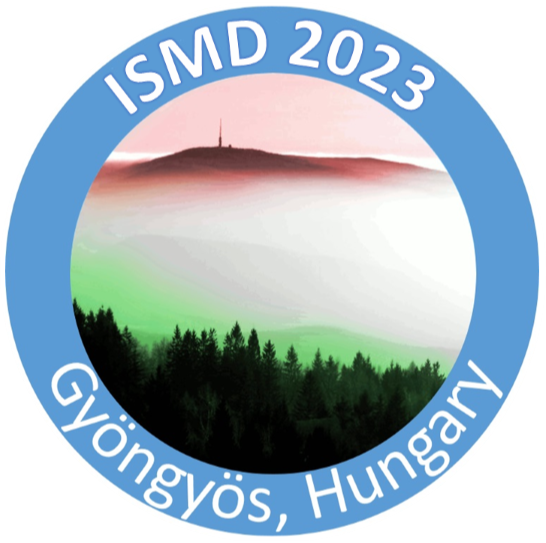}\\[1cm]
	\textbf{Nuclear modification factors and the Cronin effect}}
\author{{Thomas A.\ Trainor$^{1}$,}\\[1ex]
	$^1$University of Washington, Seattle, USA\\
}

\begin{document}

\maketitle

\begin{abstract} 
Nuclear modification factors (NMFs) applied to A-B collision systems consist of $p_t$ spectrum ratios rescaled by an estimated number of nucleon-nucleon binary collisions. Interest in NMFs is motivated by possible modification (suppression?) of jet production in more-central A-B collisions conjectured to arise from a deconfined quark-gluon plasma or QGP. Interpretation of NMFs is complicated by the so-called Cronin effect wherein similar ratios derived from fixed-target p-A data exhibited suppression at lower $p_t$ and enhancement at higher $p_t$ with increasing atomic weight A. This presentation describes precision analysis of identified-hadron spectra from 5 TeV $p$-Pb collisions that accurately isolates the entire jet contribution. Evolution of NMF spectrum ratios with $p$-Pb centrality is interpreted in terms of variation of corresponding jet contributions to spectra. The same method is then applied to Chicago-Princeton fixed-target spectrum data from the seventies wherein the Cronin effect was first observed. Soft and hard particle densities vary as fixed powers of $A$. Inferred jet contributions are quantitatively consistent with extrapolation from higher energies. The Cronin effect is a simple result of rescaling particle-density spectra by factor $1/A^{1/3}$.

\end{abstract}

\section{Introduction}

This presentation summarizes a study as presented in Ref.~\cite{tomnmf} of nuclear modification factors (NMFs) and their relation to the so-called Cronin effect first observed by the Chicago-Princeton (C-P) collaboration. Concerning NMFs, how is rescaling by \nn\ binary-collision number $N_{bin}$ determined, how are NMFs physically interpreted and how are NMFs related to LHC jets? Concerning the Cronin effect, how does the observed effect relate to details of C-P spectrum data and how are C-P spectra related to LHC jets?

\section{A two-component model (TCM) for PID spectra}

Given a \pt\ spectrum TCM for unidentified-hadron spectra~\cite{ppprd,alicetomspec,newpptcm} a corresponding TCM for identified hadrons (i.e.\ PID) can be generated by assuming that each hadron species $i$ comprises certain {\em fractions} of soft and hard TCM components denoted by $z_{si}(n_s)$ and $z_{hi}(n_s)$  with $z_{xi}(n_s) \leq 1$. A PID spectrum TCM can then be expressed as
\bea \label{pidspectcm}
\bar \rho_{0i}(p_t,n_s) &\approx&  d^2 \bar n_{chi} / p_t dp_t dy_z
\\ \nonumber
&=& S_i(p_t) + H_i(p_t,n_s)
\\ \nonumber
&=& z_{si}(n_s) \bar \rho_{s} \hat S_{0i}(p_t) + z_{hi}(n_s) \bar \rho_{h} \hat H_{0i}(p_t,n_s),
\eea
where $\bar \rho_0 = \bar \rho_{s} + \bar \rho_h$ is the measured event-class total particle density, and particle densities averaged over pseudorapidity angular acceptance $\Delta \eta$ are defined by $\bar \rho_x = n_x / \Delta \eta$. Factorization of soft and hard components separately into a particle density and a unit-normal \pt\ function, completely (soft) or nearly (hard) independent of particle density or A-B centrality (for small collision systems), is a unique feature of the TCM as inferred from data~\cite{ppprd}.  For \pp\ collisions $\bar \rho_h \approx \alpha(\sqrt{s}) \bar \rho_s^2$~\cite{alicetomspec,newpptcm} and $\bar \rho_s$ is then obtained from measured  $\bar \rho_0$ as the root of $\bar \rho_0 = \bar \rho_{s} + \alpha \bar \rho_s^2$. For \ppb\ collisions inference of $\bar \rho_s$ requires more-elaborate centrality determination~\cite{ppbpid}. \ppb\ data spectra are plotted here as densities on \pt\ (as published) vs transverse rapidity $y_{t} = \ln[(p_t + m_{t\pi})/m_\pi]$.  Unit-normal model functions $\hat S_{0i}(m_t)$ (soft) and $\hat H_{0i}(y_t,n_s)$ (hard) are defined on those respective variables where they have simple forms (exponential on \mt\ with power law tail and Gaussian on \yt\ with exponential tail respectively), and are then transformed to \pt\ via appropriate Jacobians as required. Detailed definitions are provided in Refs.~\cite{ppbpid,pidpart1,pidpart2}. For convenience note that \yt\ = 2 corresponds to $p_t \approx 0.5$ GeV/c, \yt\ =2.7 to 1 GeV/c, \yt\ =4 to 3.8 GeV/c and \yt\ = 5 to 10 GeV/c.

\section{5 TeV p-Pb PID spectrum data}

The 5 TeV \ppb\ PID spectrum data shown below were reported by Ref.~\cite{alicenucmod} as an extension to higher \pt\ of PID spectra from the same collision system reported in Ref.~\cite{aliceppbpid}.
Collision events are sorted into seven multiplicity classes. Mean charge densities $\langle dN_{ch} / d\eta \rangle \rightarrow \bar \rho_0$  as integrated within $|\eta| < 0.5$ or angular acceptance $\Delta \eta = 1$ are 45, 36.2, 30.5, 23.2, 16.1, 9.8 and 4.4 for event classes $n \in [1,7]$ ($n=1$ is most central)~\cite{aliceppbpid}. 

\begin{figure}[h]
	\includegraphics[width=1.46in]{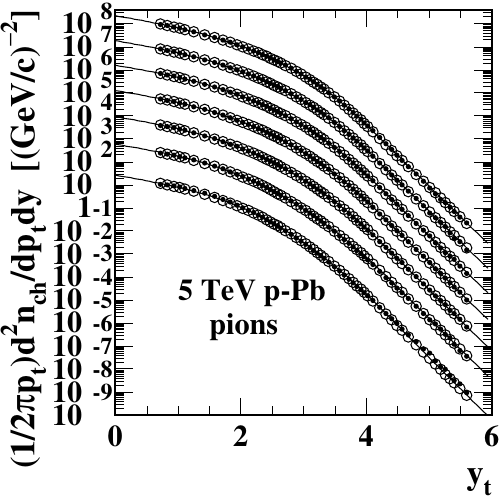}
	\includegraphics[width=1.44in]{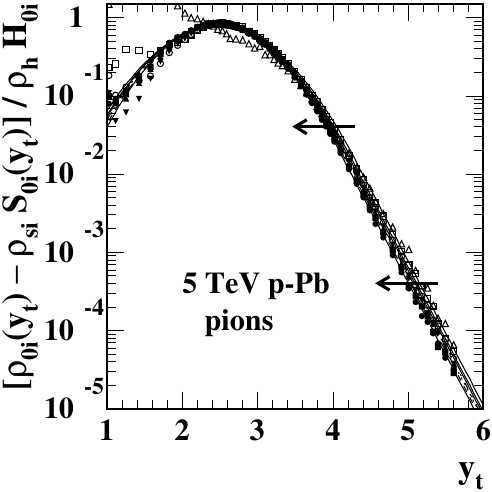}
	\includegraphics[width=1.46in]{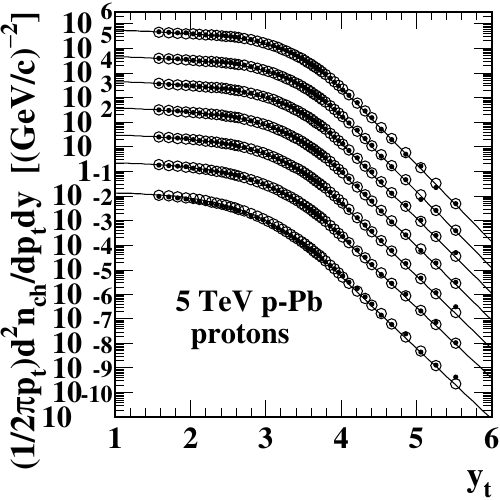}
	\includegraphics[width=1.44in]{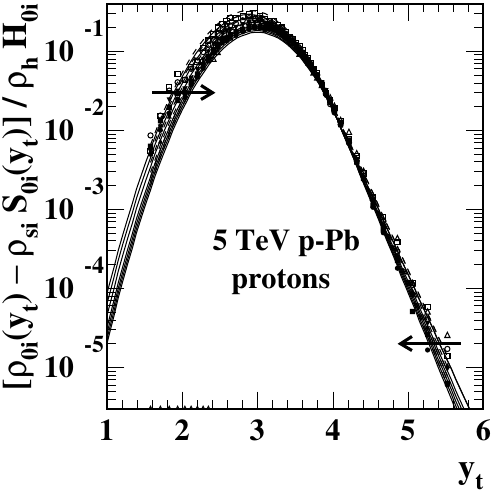}
\put(-343,90) {\bf (a)}
	\put(-237,90) {\bf (b)}
	\put(-128,90) {\bf (c)}
	\put(-23,90) {\bf (d)}
	\caption{\label{fig1}
		(a,c) \pt\ spectra from 5 TeV \ppb\ collisions (solid dots) for pions (a) and protons (c) from Ref.~\cite{alicenucmod}. Corresponding TCM is solid curves and open circles.
		(b,d) Spectrum data hard components (points of several styles) and corresponding TCM (solid curves). The arrows indicate hard-component trends corresponding to increasing centrality.
	}   
\end{figure}

Figure~\ref{fig1} (a,c) shows PID \pt\ spectra (solid dots) for charged pions and protons plotted vs pion rapidity \yt. The solid curves (continuum) and open circles (defined on data \yt\ values) are {\em variable}-TCM spectra wherein certain parameters for the hard component vary linearly with the TCM hard/soft ratio $x(n_s)\nu(n_s)$. Results for charged kaons from Ref.~\cite{tomnmf} are here omitted for brevity. Except for event class $n = 7$ the \ppb\ data spectra are described accurately within statistical uncertainties for \pt\ from zero to 20 GeV/c.

Experience derived from PID spectrum analysis of a broad array of A-B collision systems and collision energies~\cite{ppbpid,pidpart1,pidpart2,pbpbpid} indicates that spectrum soft components modeled by $\hat S_{0i}(m_t)$ are independent of event \nch\ or A-B centrality, permitting accurate isolation of complementary jet-related spectrum hard components by a simple procedure as follows: Given the structure of Eq.~(\ref{pidspectcm}) spectrum {\em data} hard components (left-hand side below) approximated by model functions $\hat H_{0i}(p_t,n_s)$ (right-hand side) are obtained by
\bea \label{pidspectcmx}
 \frac{\bar \rho_{0i}(p_t,n_s) - z_{si}(n_s)\bar \rho_s \hat S_{0i}(p_t)}{ \hat H_{0i}(\bar p_t,n_s)\bar \rho_h} &\approx& z_{hi}(n_s) \frac{\hat H_{0i}(p_t,n_s)}{\hat H_{0i}(\bar p_t,n_s)},
\eea
where $\hat H_{0i}(\bar p_t,n_s)$ is the value of model $\hat H_{0i}(p_t,n_s)$ at its mode. The choice of rescaling at left generates data distributions whose maxima approximate hard-component fractions $z_{hi}(n_s)$ as implied by the right hand side of Eq.~(\ref{pidspectcmx}).
The accuracy of such differential analysis relies on precise determination of \ppb\ geometry, nonPID particle densities $ \bar \rho_{s}$ and $ \bar \rho_{h}$ and PID species fractions $z_{si}(n_s)$ and $z_{hi}(n_s)$~\cite{ppbpid,tomglauber,tomexclude}.

Figure~\ref{fig1} (b,d) shows spectrum hard components in the format of Eq.~(\ref{pidspectcmx}). It is notable that in Eq.~(\ref{pidspectcmx}) soft-component model $\hat S_{0i}(p_t)$ has only one adjustable parameter, exponent $n$, that is constrained by a well-defined collision-energy trend over three orders of magnitude (see Fig.~\ref{fig5} (b) below). Coefficients $z_{si}(n_s)\bar \rho_s$ derived from earlier \ppb\ studies must correspond to data at the lowest \pt\ values, which is the case. PID spectrum hard components (approximately 100\% of the jet fragment distributions) are thereby isolated and described by the TCM accurately to well below 0.5 GeV/c ($y_t \approx 2$). Figure~\ref{fig1} (b,d) provides the most complete and differential description of jet contributions to PID spectra possible, retaining  all relevant information carried by particle data (particle momenta and hadron species). The TCM procedure is an example of {\em data compression} -- from terabytes of individual particle momentum data for a typical collision system to tens of bytes of TCM parameter values {\em that are not derived from model fits to individual spectra}.

\section{Nuclear modification factors -- NMFs} \label{sec4}

Nuclear modification factors as conventionally utilized are spectrum ratios, e.g.\ A-B spectrum in ratio to \pp\ spectrum, rescaled by an estimate of the average number of \nn\ binary collisions $N_{bin}$ within A-B collisions of a given centrality. For A-B $\rightarrow$ \ppb\ the definition $R_{pPb} \equiv (1/N_{bin}) \bar \rho_{pPb}(p_t) / \bar \rho_{pp}(p_t)$ is based on the assumption that, other things being equal, \nn\ collisions within A-B collisions are on average equivalent to NSD \pp\ collisions. Estimation of $N_{bin}$ is in effect based on the same assumption (see below). If  there are no significant jet modifications then application of NMFs to {\em any} \ppb\ centrality should produce the same result, with $R_{pPb} \approx 1$ at higher \pt. The trend at lower \pt\ tends to be ignored. In what follows two issues are considered: (a) how accurate are $N_{bin}$ estimates and (b) how should unrescaled spectrum ratios ($R'_{pPb}$ below) behave per the TCM.

\subsection{A-B centrality and number of binary collisions $\bf N_{bin}$}

Estimation of $N_{bin}$ for given A-B collision system requires determination of collision centrality vs a measured quantity such as particle density $\bar \rho_0$ at midrapidity. Conventionally, the centrality problem is addressed via a classical Glauber Monte Carlo. For 5 TeV \ppb\ collisions an example is given by Ref.~\cite{aliceglauber} wherein certain assumptions are made: ``Under the assumption that the multiplicity measured in the Pb-going rapidity region scales with the number of Pb-participants, an approximate independence of the multiplicity per participating nucleon measured at midrapidity of the number of participating nucleons is observed.'' In simpler language particle density at midrapidity is assumed to be proportional to $N_{part}$ for any $N_{part}$. There is also the conclusion ``...at high-\pt\ the \ppb\ spectra are found to be consistent with the pp spectra scaled by $N_{coll}$ [$N_{bin}$] for all centrality classes'' which translates to $R_{pPb} \approx 1$ at higher \pt. But particle density $\bar \rho_0$ at midrapidity includes a contribution from jets that manifest {\em in part} at high \pt\ but {\em most abundantly} near 1 GeV/c (see Fig.~\ref{fig1}, b,d) where they contribute substantially to particle density $\bar \rho_0$ at midrapidity. There is thus a basic contradiction among those assumptions. The ALICE study concludes that ``...nuclear modification factors are consistent with unity above $\sim 8$ GeV/c, with no centrality dependence[, and] that the multiplicity of charged particles at mid-rapidity scales linearly with the total number of participants....'' That scenario is dramatically inconsistent with ALICE ensemble mean \mmpt\ measurements in Ref.~\cite{alicempt} as demonstrated in Refs.~\cite{ppbpid,alicetommpt,tommpt} where the correct centrality determination for 5 TeV \ppb\ collisions is established. A larger context for the latter analysis is presented in Refs.~\cite{tomglauber,tomexclude}.

\subsection{Spectrum ratio trends}

Equation~(\ref{rpbp}) defines spectrum ratios $R_{p\text{Pb}i}'$ expressed in terms of a PID TCM for \ppb\ collisions. The prime indicates that estimated $1/N_{bin}$ has been omitted from conventional NMF $R_{p\text{Pb}}$. The first line is from Eq.~(\ref{pidspectcm}) that describes individual PID spectra within their point-to-point data uncertainties, with $\bar \rho_s = (N_{part}/2) \bar \rho_{spN}$ and $\bar \rho_h = N_{bin} \bar \rho_{hpN}$:
\bea \label{rpbp}
R_{p\text{Pb}i}' &=& \frac{ z_{sipPb}(n_s) \bar \rho_{s} \hat S_{0ipPb}(p_t) +   z_{hipPb}(n_s) \bar \rho_{h} \hat H_{0ip\text{Pb}}(p_t,n_s)}{z_{sipp} \bar \rho_{spp} \hat S_{0ipp}(p_t) + z_{hipp} \bar \rho_{hpp} \hat H_{0ipp}(p_t)}~~
\\ \nonumber 
&\rightarrow& \frac{ z_{sipPb}(n_s) (N_{part}/2) \bar \rho_{spN} \hat S_{0ipPb}(p_t)}{z_{sipp} \bar \rho_{spp} \hat S_{0ipp}(p_t)} ~~\text{for low \pt}
\\ \nonumber 
&\rightarrow& \frac{z_{hipPb}(n_s) N_{bin} \bar \rho_{hpN} \hat H_{0ip\text{Pb}}(p_t,n_s)}{z_{hipp} \bar \rho_{hpp} \hat H_{0ipp}(p_t)}~~\text{for high \pt}.
\eea
For low \pt, soft-component model $\hat S_{0i}(p_t)$ is independent of multiplicity or centrality for any collision system. Refer to Fig.~\ref{fig1} where spectra maintain the same form within statistical uncertainties below \yt\ = 2 ($p_t \approx 0.5$ GeV/c) and where assumption of a fixed $\hat S_{0i}(p_t)$ model function leads to isolation of spectrum hard components with intact shapes down to or even below 0.5 GeV/c. If $\hat S_{0i}(p_t)$ cancels in ratio what remain are factors $z_{si}(n_s)\bar \rho_{sNN}$ for \ppb\ and \pp\ collisions and geometry parameter $N_{part}/2$ {\em that should be common to all hadron species}. If \ppb\ collisions are linear superpositions of inelastic \pn\ collisions, as assumed by ALICE, cancellation of $z_{si}(n_s)\bar \rho_{sNN}$ factors leaves $N_{part}/2$ in isolation implying that $R_{p\text{Pb}}'$ at low \pt\ provides an estimate {\em from spectrum data} of $N_{bin} = N_{part} - 1$ for \ppb.

For high \pt, and based on the same linear-superposition assumption, factors $z_{hi}(n_s)\bar \rho_{hNN}$ from \ppb\ and \pp\ collisions should cancel leaving the expression $N_{bin}\hat H_{0ip\text{Pb}}(p_t,n_s)/\hat H_{0ipp}(p_t)$. In  the absence of jet modification that expression simplifies to $N_{bin}$ alone (in principle determined from the low-\pt\ limiting case) justifying the definition of conventional {\em unprimed} $R_{p\text{Pb}}$ based on several strong assumptions as noted.

\section{NMF ratio data}

\begin{figure}[h]
	\includegraphics[width=2.9in]{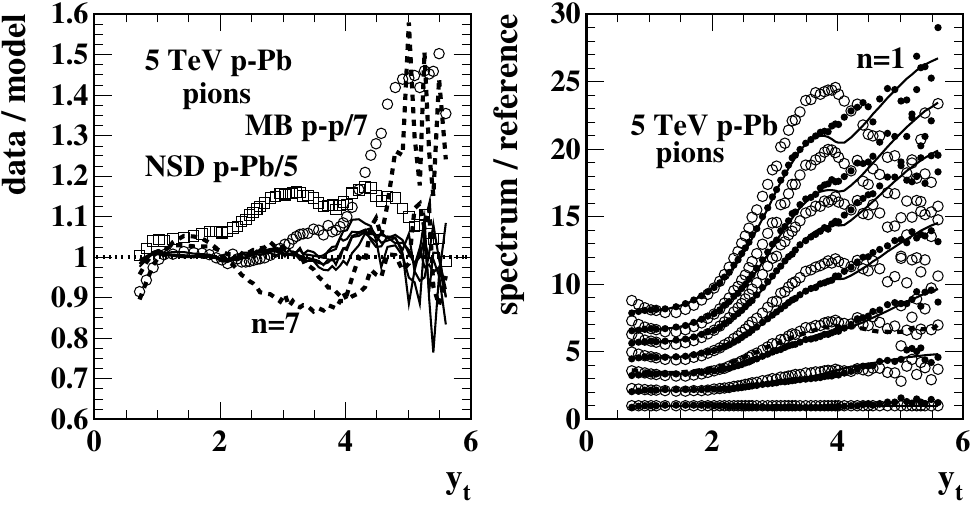}
	\includegraphics[width=2.9in]{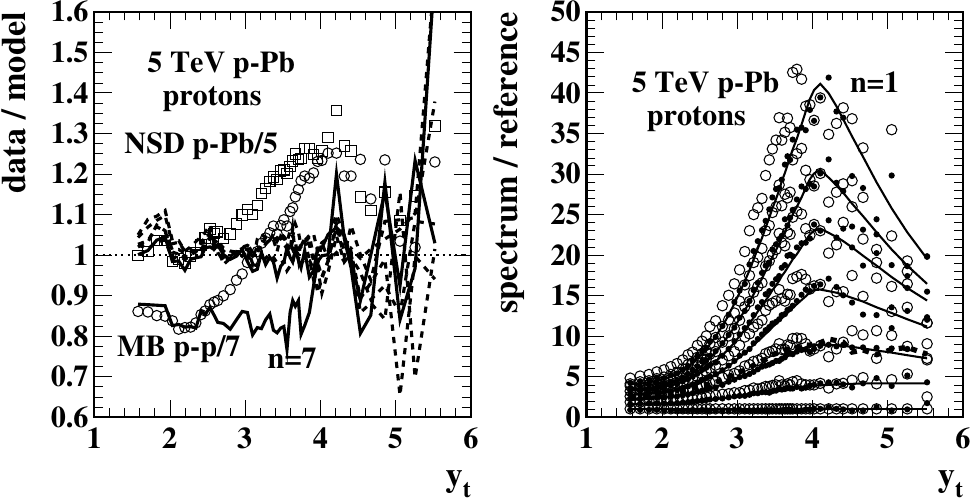}
\put(-353,93) {\bf (a)}
\put(-257,93) {\bf (b)}
\put(-135,93) {\bf (c)}
\put(-23,93) {\bf (d)}
\caption{\label{fig2}
	(a,c) Ratios of data spectra to TCM for seven event classes (dashed and solid curves) and ratios of NSD \ppb\ to $n=5$ TCM (open squares) and inelastic \pp\ to $n=7$ TCM (open circles).
	(b,d) Ratios to TCM $n = 7$ (peripheral) of data (solid dots) and corresponding TCM (solid curves) and ratios of data spectra to data $n = 7$ (open circles).
	}   
\end{figure}

Figure~\ref{fig2} (a,c) shows ratios of spectrum data to TCM (solid and dashed curves) for seven event classes from 5 TeV \ppb\ collisions. The data/TCM ratios for most-peripheral events ($n = 7$) are strongly biased as is typically observed across collision systems. Also shown are non-single-diffractive (NSD average) \ppb\ spectra in ratio to TCM $n = 5$ (open squares) and inelastic (MB) \pp\ spectra in ratio to TCM $n = 7$ (open circles).

Figure~\ref{fig2} (b,d) shows data (solid dots) and TCM (solid curves) in ratio to TCM $n = 7$ as reference for seven event classes. Also shown are data in ratio to {\em data} $n =7$ as reference (open circles) to demonstrate the effect of the biased $n = 7$ data spectra. Just visible are NSD \ppb\ spectra in ratio to inelastic \pp\ spectra (dashed curves).

These results reveal that the ALICE linear-superposition assumption fails dramatically. The low-\pt\ limits for $n = 1$ (most central) are approximately 8, 6 and 4 for pions, charged kaons (not shown) and protons respectively whereas one should expect a single value common to all hadron species for linear superposition. Also, given $N_{part} = N_{bin} + 1$ for \ppb, at high \pt\ one would expect $N_{bin} \approx 15$, 11 and 7 for the three species respectively, but the high-\pt\ values for data and TCM are 23 and 40 for pions and protons respectively. Thus, direct evidence from spectrum ratios demonstrates that the basic linear-superposition assumption underlying ALICE \ppb\ centrality determination in Ref.~\cite{aliceglauber} is incorrect.

\section{Variable vs fixed hard-component $\bf H_0(p_t)$ models}

\begin{figure}[h]
	\includegraphics[width=1.45in]{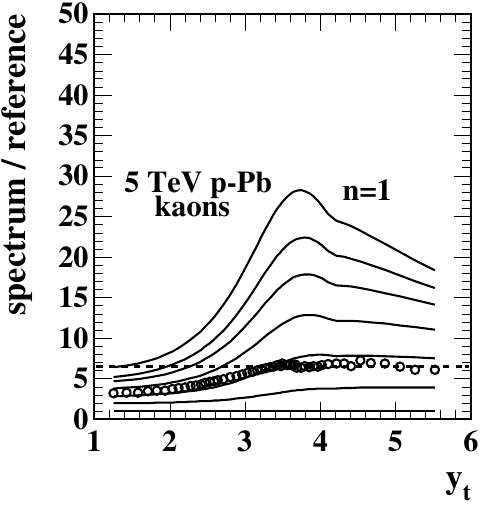}
	\includegraphics[width=1.45in]{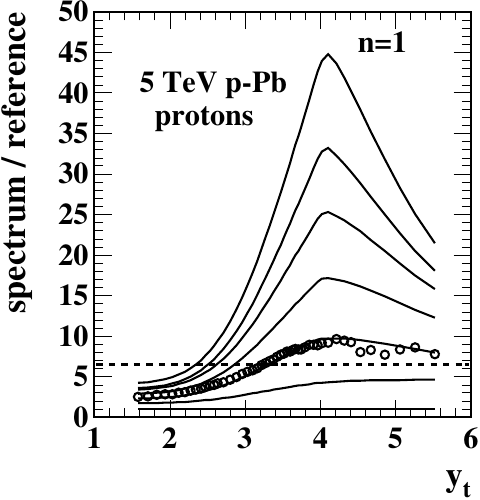}
	\includegraphics[width=1.45in]{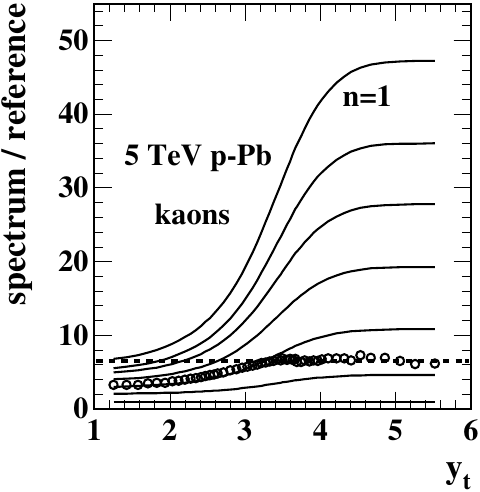}
	\includegraphics[width=1.45in]{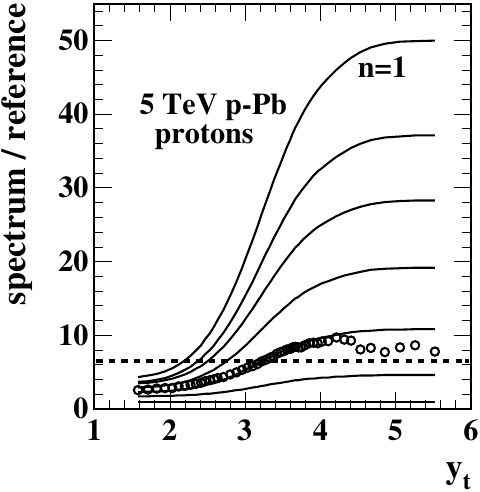}
\put(-345,87) {\bf (a)}
\put(-239,87) {\bf (b)}
\put(-158,93) {\bf (c)}
\put(-53,93) {\bf (d)}
\caption{\label{fig3}
	(a,b) TCM ratios to TCM $n = 7$ as in Fig.~\ref{fig2} (d) for kaons (a) and protons (b). The open circles are data NSD \ppb\ in ratio to inelastic \pp.
	(c,d) Curves as in (a,b) except that hard components are held fixed independent of \ppb\ centrality and thus cancel.
	}  
\end{figure}

Figure~\ref{fig3} (a,b) shows TCM spectrum ratios (solid curves) for charged kaons (a) and protons (b) with data points omitted for clarity. Also plotted are NSD \ppb\ spectra in ratio to inelastic \pp\ spectra (open circles) as reported in Ref.~\cite{alicenucmod}. Those ratios correspond approximately to TCM ratios for data event class $n = 5$ in ratio to TCM $n = 7$. The ALICE estimate $N_{bin} \approx 6.5$ for those data is represented by the dashed  lines, which implies a value $R_{pPb} \approx 1$ as expected for no nuclear modification of jets. However,  the correct $n = 5$ value is $N_{bin} \approx 1.3$ as presented in Table 2 of Ref.~\cite{ppbpid}. It should be noted that {\em measured} $\bar \rho_0$ for that event class is 16.1 at which point ensemble \mmpt\ for \ppb\ is indistinguishable from \mmpt\ for \pp\ collisions {\em with the same} $\bar \rho_0$. See Fig.~3 (left) of Ref.~\cite{ppbpid}.  That is, most \ppb\ collisions in event class $n = 5$ are single peripheral \pn\ collisions. \ppb\ centrality determination is directly related to the magnitude  of jet production as revealed by \mmpt\ trends in Refs.~\cite{alicetommpt,tommpt}.

Figure~\ref{fig3} (c,d) shows the TCM ratios in panels (a,b) but with hard-component models $\hat H_{0ixy}(p_t)$ held fixed, in which case they cancel at higher \pt\ in Eq.~(\ref{rpbp}). What remains above $y_t \approx 4$ is the ratio $N_{bin} z_{hipPb} \bar \rho_{hipN} / z_{hipp} \bar \rho_{hipp}$. The $z_{hi}$ ratios are $O(1)$. Consider the $n=1$ most-central event class. From Table 2 of Ref.~\cite{ppbpid} $N_{bin} \approx 3.2$ for that event class. How then explain the limiting value $\approx 50$ in Fig.~\ref{fig3} (d)? From the same Table 2 $\bar \rho_{sNN} \approx 4.2$ for peripheral $n = 7$ $\approx$ \pp\ and 16.6 for central $n = 1$ \ppb, i.e. a factor 4 increase. But $\bar \rho_h \propto \bar \rho_s^2$ for elementary \nn\ collisions within a linear system (i.e.\ no QGP). Thus, the $\bar \rho_h$ ratio in the above expression introduces a {\em factor 16} leading to $3.2 \times 16 \approx 50$ in Fig.~\ref{fig3} (d) at high \pt. The large high-\pt\ ratio values, accurately consistent with spectrum data from Ref.~\cite{alicenucmod}, arise from variation of \pn\ properties with increasing \ppb\ centrality that is dramatically inconsistent with assumed linear superposition. It is also inconsistent with the conclusion from Ref.~\cite{alicenucmod} that ``at high-\pt\ the \ppb\ spectra are found to be consistent with the pp spectra scaled by [$N_{bin}$] [i.e.\ NMF $\approx 1$] {\em for all centrality classes} [emphasis added].'' See the ratio trends in Fig.~\ref{fig2} (b,d) where, for more-central collisions, no fixed value of $N_{bin}$ could produce data NMFs consistent with unity at high \pt. In fact only one ratio, NSD \ppb\ in ratio to inelastic \pp\ (open circles above), is reported in Ref.~\cite{alicenucmod}.

\section{Species/species spectrum ratios}

\begin{figure}[h]
	\includegraphics[width=2.885in]{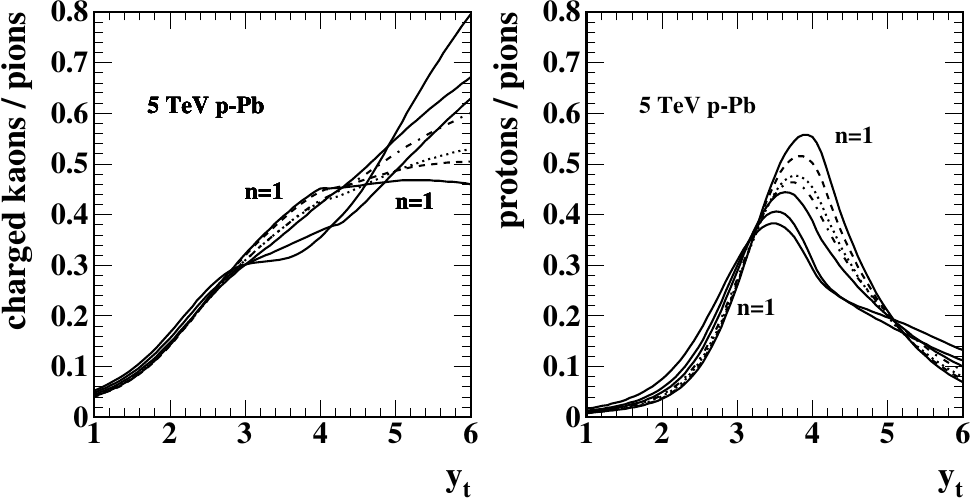}
	\includegraphics[width=2.915in]{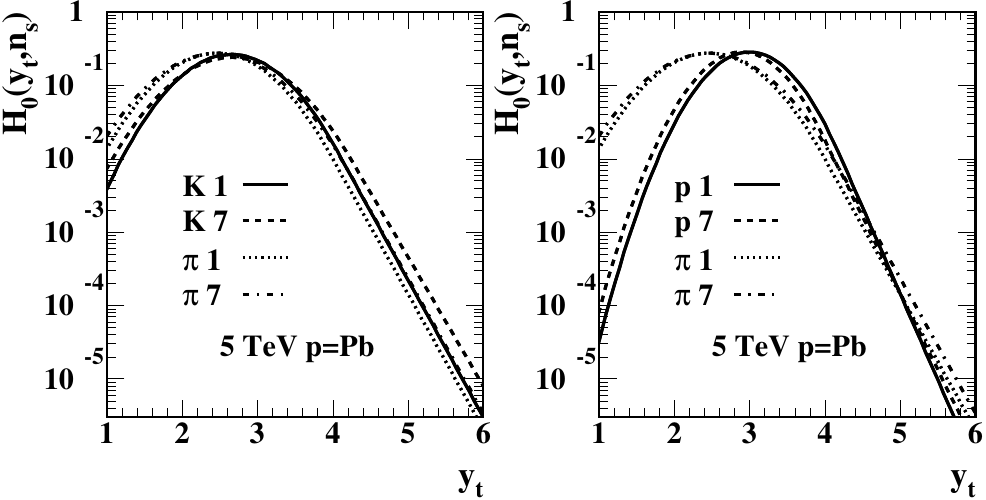}
\put(-353,90) {\bf (a)}
\put(-239,90) {\bf (b)}
\put(-128,90) {\bf (c)}
\put(-23,90) {\bf (d)}
\caption{\label{fig4}
	Species/species ratios for kaons/pions (a) and protons/pions (b).
	Comparison of TCM  hard components for pions vs kaons (c) and pions vs protons (d).
	}  
\end{figure}

Figure~\ref{fig4} (a,b) shows TCM charged-kaon/pion and proton/pion (species/species) spectrum ratios respectively as another application of ratios to investigation of high-energy nuclear collisions. Emphasizing here the proton/pion ratio, the prominent peak near $y_t = 3.8$ ($p_t \approx 3$ GeV/c) with its centrality evolution has been associated with radial flow \cite{hydro} or quark recombination~\cite{reco}.
Reference~\cite{hydro} makes the following observations about species/species ratios in \pp\ and \pbpb\ collisions: ``The Krakow hydrodynamical model captures the rise of both [proton/pion and kaon/pion] ratios quantitatively well....'' ``The EPOS event generator which has both hydrodynamics, but also the high \pt\ physics and special hadronization processes for quenched jets qualitatively well describes the data....''  ``...the model comparisons [noted above]...indicate that the [proton/pion] peak [near 3 GeV/c] is mainly dominated by radial flow (the masses of the hadrons).'' Reference~\cite{alicenucmod}, commenting on 5 TeV \ppb\ PID spectra, states that ``for \pt\ below 2-3 GeV/c the spectra behave like in \pbpb\ collisions, i.e., the \pt\ distributions become harder as the multiplicity increases and the change is most pronounced for protons and lambdas. In heavy-ion collisions this effect is commonly attributed to radial flow.''  But as demonstrated in panels (c,d) below that \pt\ interval ($y_t < 3.7$) includes most of the jet fragments~\cite{jetspec2,mbdijets}.

Figure~\ref{fig4} (c,d) shows TCM hard components taken (for pions and protons) from Fig.~\ref{fig1} (b,d). Line styles distinguish most-peripheral ($n = 7$) and most-central ($n = 1$) curves. Comparing panel (b) with panel (d) it should be apparent that the peaked structure in the proton/pion ratio is due to the interplay between corresponding hard-component (jet fragment) shapes, the proton hard component being narrower and with mode at substantially higher \yt\ consistent with {\em measured} fragmentation functions~\cite{ppbpid,eeprd,fragevo}. This comparison demonstrates that the structure of species/species spectrum ratios is dominated above \yt\ = 2.5 ($p_t \approx 1$ GeV/c) by minimum-bias jet fragment contributions. Note that if radial flow were responsible for the proton/pion peak one would have to explain the shift of pion (and kaon) hard components to {\em smaller \yt} with increasing \ppb\ centrality as in Fig.~\ref{fig1} (b).

\section{Chicago-Princeton (C-P) PID spectra}

One source of confusion regarding NMFs is the {\em Cronin effect} in which NMFs formed from \pa/\pp\ spectrum ratios may exhibit a localized increase relative to unity near and above 1 GeV/c reducing to unity at higher \pt\ with a decrease below unity or suppression at low \pt~\cite{boris}. The Cronin effect was first observed in the seventies by the Chicago-Princeton (C-P) collaboration operating at the National Accelerator Laboratory, with J. Cronin as a  member~\cite{cronin10,cronin0}. Spectrum data were published as invariant differential {\em cross sections} in the form $Ed\sigma_{pA}/d^3p \rightarrow \sigma_{pA}d^2n_i / 2\pi m_t dm_t dy_z$, where $\sigma_{pA}$ (in units of $cm^2$) is an absorption total cross section and $n_i$ is a yield for hadron species $i$.  In what follows published spectrum data are multiplied by common factor $2\pi 10^{27} / \sigma_{pA}(mb)$ to yield particle densities on \mt\ and \yz\ of the form $d^2n_i / m_t dm_t dy_z$. The TCM is then applied to C-P data in that form.

\subsection{TCM soft and hard model functions vs collision energy} \label{tcmmodel}

The energy dependences of TCM model functions $\hat S_0(m_t)$ and $\hat H_0(y_t)$ for C-P data are determined from \pp\ and \pw\ spectra respectively at three collision energies~\cite{cronin10,cronin0} corresponding to proton beam energies of 200, 300 and 400 GeV as described below.

\begin{figure}[h]
	\includegraphics[width=1.48in]{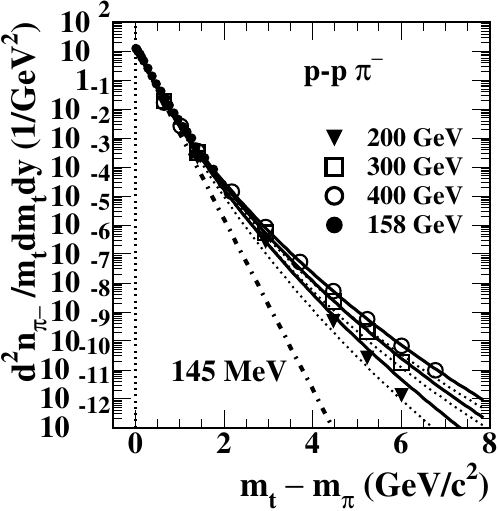}
	\includegraphics[width=1.42in]{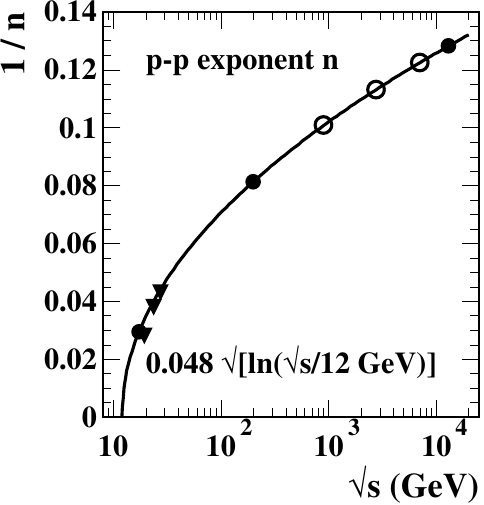}
	\includegraphics[width=1.48in]{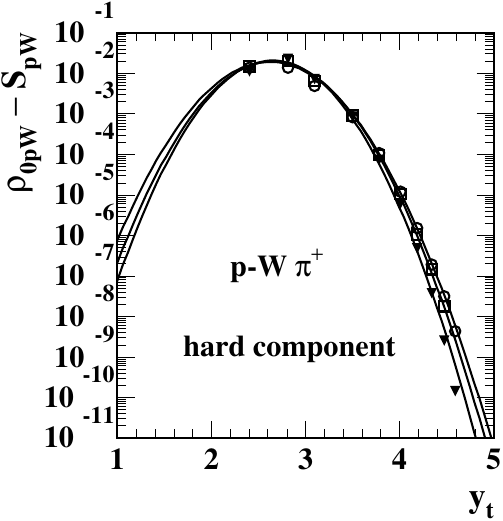}
	\includegraphics[width=1.42in]{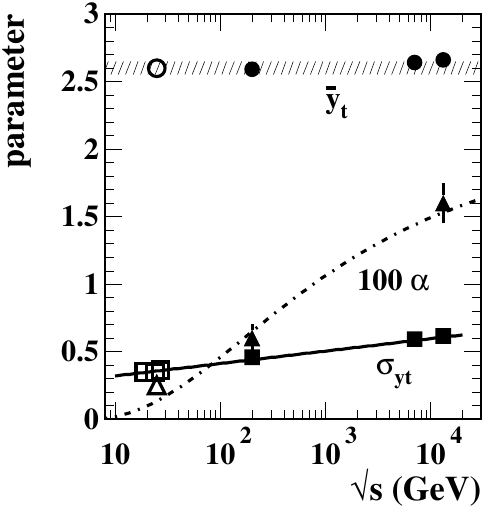}
\put(-343,90) {\bf (a)}
\put(-237,75) {\bf (b)}
\put(-128,90) {\bf (c)}
\put(-23,80) {\bf (d)}
\caption{\label{fig5}
	(a) C-P pion spectra for three beam energies (points) with TCM soft components (dotted) and full TCM spectra (solid).
	(b) Collision energy trend for TCM soft-component exponent $n$ at C-P energies (triangles) compared to previously-published results from higher energies (other points). The curve is explained in the text.
	(c) C-P pion hard components (points) and corresponding TCM models (curves).
	(d) TCM hard-component parameters $\bar y_t$, $\sigma_{y_t}$ and $\alpha$ for C-P data (open points) and for higher energies (solid points).
	} 

\end{figure}

Figure~\ref{fig5} (a) shows C-P pion spectra (points) from \pp\ collisions for three beam energies that are used to determine soft-component exponents $n$. Exponent $n$ is adjusted for each energy so that the soft component (dotted curves) lies just below the data point at highest \pt\ after the curve is rescaled to matched the point at lowest \pt. \pp\ data are chosen because hard components are minimal for that system. Figure~\ref{fig5} (b) shows the results plotted as $1/n$ (inverted triangles). Other plot features were previously published as Fig.~17 (left) in Ref.~\cite{ppbpid}, demonstrating that the C-P soft component is consistent with results from LHC energies. The solid curve is motivated by association with Gribov diffusion within a parton splitting cascade during projectile-nucleon dissociation~\cite{tomnmf,gribov,gribov2}.

Figure~\ref{fig5} (c) shows C-P pion spectrum hard components (points) from \pw\ collisions for three beam energies that are used to determine hard-component widths $\sigma_{y_t}$. Data hard components (points) are inferred by subtracting soft-component models consistent with panel (a) from published \pw\ spectra chosen because hard components are maximal for that system. Figure~\ref{fig5} (d) shows inferred widths plotted as open squares. The original plot was published as Fig.~23 (left) in Ref.~\cite{ppbpid}, also demonstrating that the TCM for C-P data is consistent with that for LHC energies as to evolution with collision energy.
 
\subsection{TCM PID particle density trends vs target size A}

Evolution of PID particle densities $\bar \rho_{si}$ and $\bar \rho_{hi}$ (i.e.\ $dn_x/dy_z$) with target size (atomic weight) $A$ is determined for fixed beam energy 400 GeV.

\begin{figure}[h]
	\includegraphics[width=2.9in]{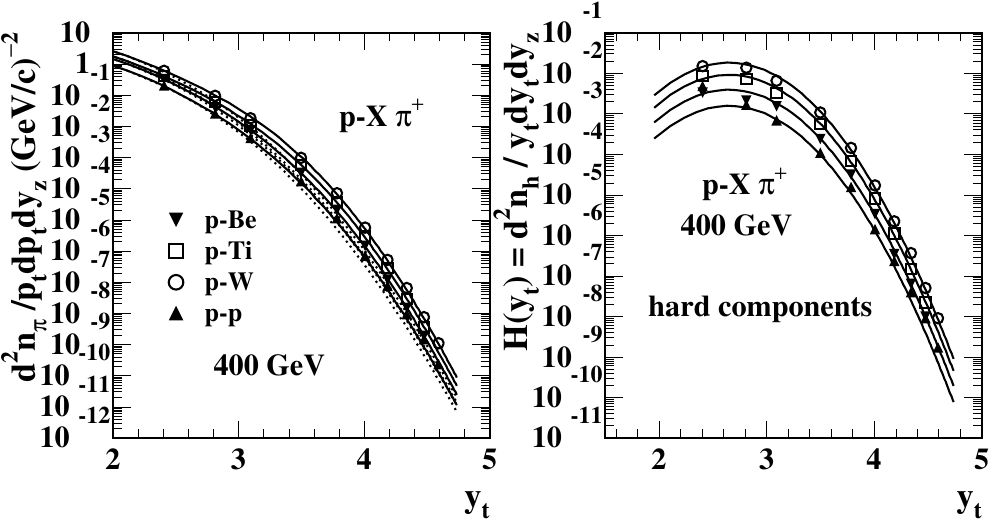}
	\includegraphics[width=1.45in]{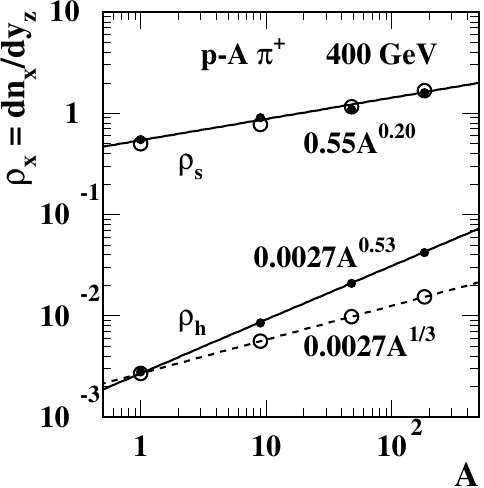}
	\includegraphics[width=1.45in]{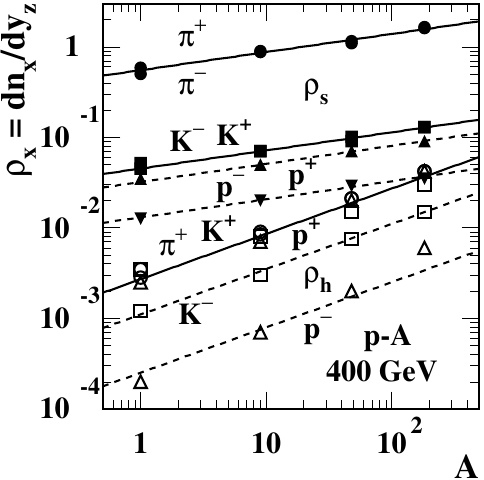}
\put(-343,90) {\bf (a)}
\put(-241,90) {\bf (b)}
\put(-128,65) {\bf (c)}
\put(-23,85) {\bf (d)}
\caption{\label{fig6}
	(a) C-P pion spectra (points) for 400 GeV beam energy and four target sizes $A$ with corresponding TCM soft components (dotted) and full TCM (solid).
	(b) C-P pion hard components (points) and corresponding TCM models (curves).
	(c) C-P soft (upper) and hard (lower) particle density trends (solid points) vs nuclear size $A$ with corresponding power-law trends (solid lines). Also shown are conjectured trends for $N_{part}$ (upper open circles) and $N_{bin}$ (lower open circles)
	(d) Similar to (c) but for all C-P hadron species.
	}  

\end{figure}

Figure~\ref{fig6} (a) shows C-P spectra for four targets $A$ (points) as densities on \pt\ plotted vs pion rapidity \yt. The dotted curves are identical soft-component shapes, derived from the 400 GeV \pp\ spectrum, rescaled to pass through the points at lowest \pt\ and thus inferring densities $\bar \rho_{si}$. Those curves are then subtracted from spectra to obtain hard-component data in panel (b). The curves are the same hard-component curve for 400 GeV inferred from \pw\ data rescaled in each case to match data and thus inferring hard densities $\bar \rho_{hi}$

Figure~\ref{fig6} (c) shows resulting $\bar \rho_{si}$ and $\bar \rho_{hi}$ trends (solid dots) vs nuclear size $A$. The solid lines indicate that the {\em individual} soft and hard trends are well described by simple power laws $A^\gamma$ with fixed values $\gamma_s \approx 0.20$ and $\gamma_h \approx 0.50$. That TCM simplicity contrasts with attempts by the C-P collaboration to describe their published intact spectra with a single complex $A^{\alpha(p_t)}$ trend~\cite{cronin0}. Note that the latter trend describes C-P {\em cross sections} while the former describes particle densities,  thus $\gamma \approx \alpha - 2/3$. The upper open circles represent a conjectured participant-number trend $N_{part} = N_{bin} + 1$ with $N_{bin} \approx A^{1/3}$ representing a path length through the target nucleus. The $\bar \rho_{si}$ data are well described by that trend. However, the same $N_{bin}$ trend applied to $\bar \rho_{hi}$ data (lower open circles, dashed line) shows a large deviation. That difference is a {\em direct manifestation of the Cronin effect}.

Figure~\ref{fig6} (d) shows that both soft and hard power-law trends are the same no matter the hadron species among pions, charged kaons and protons of both signs. It is remarkable that for positive hadrons the hard-component yield trends (jet fragments) are approximately equal while the soft-component trends may differ by an order of magnitude or more.

\section{NMF ratios vs the Cronin effect}

 NMFs formulated for C-P collaboration PID spectrum data may be defined by~\cite{accardi}
\bea \label{rab}
R_{AB} &=& \frac{B}{A} \frac{Ed\sigma_{pA} / d^3p}{Ed\sigma_{pB} / d^3p},
\eea
where each differential cross-section spectrum in numerator or denominator includes absorption total cross section $\sigma_{pX}$ as a factor.

\begin{figure}[h]
	\includegraphics[width=2.9in]{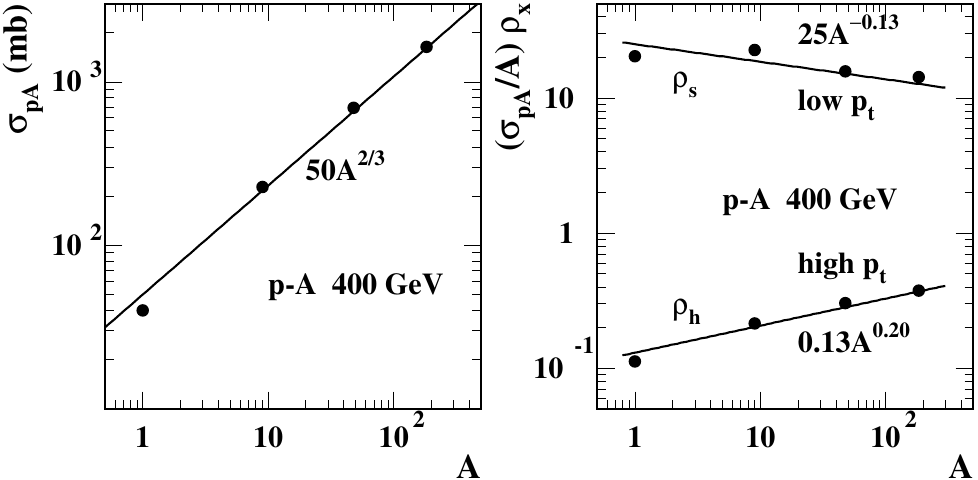}
	\includegraphics[width=2.9in]{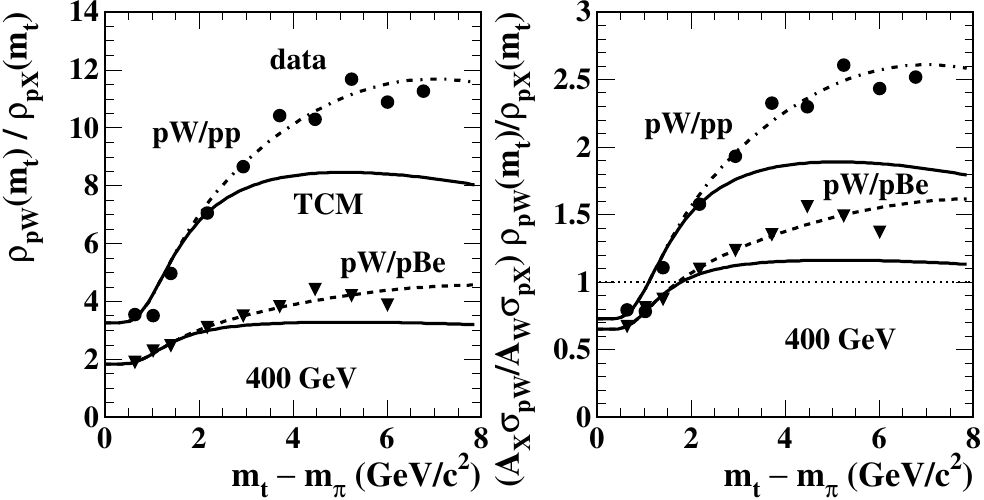}
\put(-340,75) {\bf (a)}
\put(-234,72) {\bf (b)}
\put(-128,74) {\bf (c)}
\put(-23,77) {\bf (d)}
\caption{\label{fig7}
	(a) Trend for absorption cross section $\sigma_{pA}$ (points) and corresponding power law (line).
	(b) C-P particle densities from Fig.~\ref{fig6} (c) rescaled by $A^{1/3} \sim N_{bin}$.
	(c) C-P spectrum ratios for \pw/\pp (dots) and \pw/\pbe\ (triangles). Solid curves are TCM ratios with hard components as in Sec.~\ref{tcmmodel}. Dashed curves are with hard-component widths varied by about 2\% to accommodate data.
	(d) Same as (c) but with each spectrum rescaled by $A^{1/3}$.
	}   
\end{figure}

Figure~\ref{fig7} (a) shows the trend of measured absorption cross sections vs nuclear size $A$ increasing as $A^{2/3}$ as might be expected for an area. The result of rescaling \pa\ differential cross sections by nuclear size $A$ is then rescaling of corresponding \pa\ {\em particle densities} by factor $1/A^{1/3} \sim 1/N_{bin}$ consistent with Sec.~\ref{sec4}. Figure~\ref{fig7} (b) shows the resulting trends for that rescaling applied to soft- and hard-component densities for data (points) and power-law trends (lines). The soft-component trend is just $\propto N_{part} / N_{bin}$ which trivially decreases with system size and provides no new information. The hard-component trend (Cronin effect) provides new information about jet production in \pa\ collisions, i.e.\ that jet production is not simply proportional to $N_{bin}$. Instead, jet production (i.e.\ $\bar \rho_{hi}$) from individual \pn\ collisions depends on the {\em projectile proton history} within nucleus $A$. This C-P trend corresponds to the ratio trends in Fig.~\ref{fig3} strongly increasing relative to $N_{bin}$.

Figure~\ref{fig7} (c) shows {\em unrescaled} data spectrum ratios (points) for \pw/\pbe\ (triangles) and \pw/\pp\ (dots). The solid curves are TCM ratios with fixed model functions as inferred in Sec.~\ref{tcmmodel}. The dashed curves are the TCM with hard-component Gaussian widths varied by about 2\% to accommodate data. Panel (d) shows rescaled ratios per Eq.~(\ref{rab}) with suppressions and enhancements relative to unity. The TCM curves indicate a constant trend below 0.5 GeV/c where the hard component (jets) makes no significant contribution.

One should note the dramatic difference between the amount of information carried by individual {\em intact} data hard components as in Fig.~\ref{fig1} (b,d) and by NMF ratios as in Fig.~\ref{fig7} (c,d). In the former case approximately 100\% of jet fragments is represented. Detailed evolution with centrality over the full \pt\ acceptance is clearly evident visually and is represented {\em quantitatively} within the TCM by linear variation of certain model parameters (e.g.\ Gaussian width and exponential tail parameters) that could be subjected to detailed theoretical analysis. In contrast, the latter case (ratios) presents limited information from only about 2\% of total jet fragments (what survives above $p_t \approx 2$ GeV/c). Very small absolute variation of data hard-component shapes leads to exaggerated variation of ratios that are beyond the means of any theory to confront meaningfully. Most of the jet contribution to spectra is overwhelmed within NMF ratios (i.e.\ concealed) by the nonjet soft component. Note that the high-\pt\ trend in panel (b) relates to the hard-component {\em amplitude at its mode} whereas the trends in panel (d) relate to variation of hard-component {\em tails at higher} \pt. There is no reason in principle that NMFs should approach unity at higher \pt. The actual trend depends on small differences between hard-component tails in two collision systems. Consider Fig.~\ref{fig7} (d) in relation to Figs.~\ref{fig6} (b) and 1 (b,d) as examples.

\section{C-P spectra in relation to LHC jets}

Various results from the current TCM C-P spectrum analysis may be combined to illustrate how jet production just above the energy threshold for that process near $\sqrt{s} = 10$ GeV may relate to jet production at the highest available collision energy -- 13 TeV.

\begin{figure}[h]
	\includegraphics[width=1.42in]{alice140bnew}
	\includegraphics[width=1.50in]{alicron20cc}
	\includegraphics[width=1.43in]{alice140cnew}
	\includegraphics[width=1.43in]{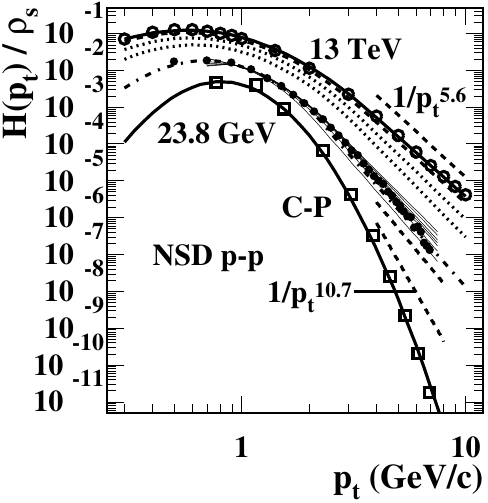}
\put(-343,80) {\bf (a)}
\put(-237,80) {\bf (b)}
\put(-128,80) {\bf (c)}
\put(-20,92) {\bf (d)}
\caption{\label{fig8}
	(a,b,c) are repeated from Fig.~\ref{fig5}.
	(d) Spectrum hard component from C-P \pp\ spectrum in the present study (open squares) and from previous NSD \pp\ studies at higher energies (other points). The solid curve is determined by C-P values (open points) in (c).
	}  
\end{figure}

Figure~\ref{fig8} (a) shows soft-component model $\hat S_0(m_t)$ exponent $n$ (defining a power-law tail extending from a Boltzmann exponential on \mt) for C-P spectra (triangles) relative to values for higher collision energies and a simple parametrization (curve) possibly related to Gribov diffusion within dissociating projectile nucleons~\cite{gribov,gribov2}. This panel illustrates that given such constraints there is little freedom within the TCM to describe spectra, i.e.\ the TCM is {\em predictive}. Soft components so defined are then used to extract hard components from published spectra as in panel (b).

Figure~\ref{fig8} (c) summarizes hard-component parameters $(\bar y_t, \sigma_{y_t})$ over the current full energy range of high-energy nuclear collisions. Again, the parameter values are simply parametrized and resulting linear trends constrain the TCM values at specific energies. Also note that the essential proportionality parameter $\alpha(\sqrt{s})$ in $\bar \rho_h \approx \alpha \bar \rho_s^2$ has its own simple parametrization (dash-dotted curve) from higher energy to which the corresponding C-P $\alpha$ value has been added (open triangle) leading to an absolute prediction for C-P hard components including magnitude as well as shape. Panel (d) shows Fig.~15 (right) from Ref.~\cite{alicetomspec} summarizing NSD \pp\ spectrum hard components from 200 GeV to 13 TeV to which has been added the equivalent from C-P data (open squares). The corresponding TCM hard component constructed from parameter values in panel (c) is the solid curve.

\section{Summary}

The subjects of this presentation are the composition, interpretation and relation to LHC jets of nuclear modification factors (NMFs) and the nature and relation to LHC jets of the Cronin effect.
Concerning NMFs, the conventional definition of spectrum ratios (e.g.\ A-B/\pp\ ratios) rescaled by an estimate of the number of binary \nn\ collisions $N_{bin}$ within A-B collisions is challenged by conventional $N_{bin}$ estimation methods and related assumptions which appear to be dramatically inconsistent with available data (e.g.\ ensemble mean \mmpt\ data). Although the intended purpose of NMFs is quantitative assessment of jet modification within A-B collisions interpretation of spectrum ratios, even before rescaling by $N_{bin}$, is also problematic since they retain very little information about a small fraction (few percent) of jet fragments. In contrast, the two-component model (TCM) of spectra quantitatively describes almost 100\% of jet fragments as a form of {\em data compression} and provides precise quantitative descriptions of evolving jet fragment distributions.

Concerning the Cronin effect, the TCM applied to fixed-target identified-hadron spectrum \pa\ data for several target masses from the Chicago-Princeton (C-P) collaboration indicates that variation of soft- and hard-component {\em shapes} with collision energy are predicted by previous TCM analyses at higher energies, consistent with basic QCD processes. C-P soft and hard {\em particle densities} vary with target size A as simple power laws. The soft trend is consistent with participant number $N_{part}$ variation. The hard trend rises substantially faster than estimated binary collision number $N_{bin} \sim A^{1/3}$. That jet-related deviation is mainly responsible for the Cronin effect as observe in NMFs: jet production from individual \pn\ collisions depends on the history of projectile protons within nuclei.

While NMFs applied to C-P data are consistent with the Cronin effect as usually described they also confirm how little information is retained by ratios. The physical origin of the Cronin effect is not apparent in ratios whereas TCM analysis establishes that the effect is due to enhancement of jet production via projectile passage through nuclei.


\end{document}